\documentclass[lettersize,journal]{IEEEtran}
\usepackage{amsmath,amsfonts}
\usepackage{algorithmic}
\usepackage{algorithm}
\usepackage{array}
\usepackage[caption=false,font=normalsize,labelfont=sf,textfont=sf]{subfig}
\usepackage{textcomp}
\usepackage{stfloats}
\usepackage{url}
\usepackage{hyperref}
\usepackage{multirow}
\usepackage{verbatim}
\usepackage{graphicx}
\usepackage{cite}
\usepackage{xcolor}
\usepackage{booktabs}
\hypersetup{
    colorlinks=true,
    linkcolor=blue,
    filecolor=blue,  
    citecolor=blue,   
    urlcolor=blue,
}
\newcommand{\modified}[1]{\textcolor{black}{#1}}

\begin{document}

\title{Efficient Image-to-Image Schrödinger Bridge for CT Field of View Extension}

\author{Zhenhao Li, {Song Ni}, Long Yang, Xiaojie Yin, Haijun Yu, Jiazhou Wang, Hongbin Han, Weigang Hu, \\Yixing Huang
\thanks{This work involved human subjects or animals in its research. The authors confirm that all human/animal subject research procedures and protocols are exempt from review board approval (since only X-ray imaging data of human subjects from public data repository or other research projects were reused in this work).}

\thanks{Z. Li, H. Yu, H. Han and Y. Huang are with Institute of Medical Technology, Peking University Health Science Center, Beijing, China. Z. Li, L. Yang, X. Yin, J. Wang and W. Hu. are with Shanghai Cancer Center, Fudan University, Shanghai, China. S. Ni is with Department of Electrical and Computer Engineering, University of Massachusetts Lowell, Lowell, USA. H. Han and Y. Huang are also with Beijing Key Laboratory of Intelligent Neuromodulation and Brain Disorder Treatment, Beijing, China. Correspondence: huangyx@pku.edu.cn (Y.H.)}
}

\markboth{Journal of \LaTeX\ Class Files,~Vol.~14, No.~8, August~2021}%
{Shell \MakeLowercase{\textit{et al.}}: A Sample Article Using IEEEtran.cls for IEEE Journals}


\maketitle

\begin{abstract}
Computed tomography (CT) is a cornerstone imaging modality for non-invasive, high-resolution visualization of internal anatomical structures. However, when the scanned object exceeds the scanner’s field of view (FOV), projection data are truncated, resulting in incomplete reconstructions and pronounced artifacts near FOV boundaries. Conventional reconstruction algorithms struggle to recover accurate anatomy from such data, limiting clinical reliability.
Deep learning approaches have been explored for FOV extension, with diffusion generative models representing the latest advances in image synthesis. Yet, conventional diffusion models are computationally demanding and slow at inference due to their iterative sampling process. To address these limitations, we propose an efficient CT FOV extension framework based on the image-to-image Schrödinger Bridge (I$^2$SB) diffusion model. Unlike traditional diffusion models that synthesize images from pure Gaussian noise, I$^2$SB learns a direct stochastic mapping between paired limited-FOV and extended-FOV images. This direct correspondence yields a more interpretable and traceable generative process, enhancing anatomical consistency and structural fidelity in reconstructions.
I$^2$SB achieves superior quantitative performance, with root-mean-square error (RMSE) values of 49.8\,HU on simulated noisy data and 152.0\,HU on real data—outperforming state-of-the-art diffusion models such as conditional denoising diffusion probabilistic models (cDDPM) and patch-based diffusion methods. Moreover, its one-step inference enables reconstruction in just 0.19\,s per 2D slice, representing over a 700-fold speedup compared to cDDPM (135\,s) and surpassing DiffusionGAN (0.58\,s), the second fastest. This combination of accuracy and efficiency {indicates that I$^2$SB has potential} for real-time or clinical deployment.

\end{abstract}


\section{Introduction}

\IEEEPARstart{C}{omputed} tomography (CT) enables high-resolution, non-invasive imaging of internal structures, with widespread applications in the medical fields. In a typical CT scanning scenario, a detector rotates around an imaged object to acquire projection data for reconstruction. However, in many clinical applications such as radiation therapy \cite{fonseca2021evaluation,gao2023transformer}, spine surgery \cite{Fan2022Fiducial}, body composition analysis \cite{XU2023102852,liman2024diffusion}, and brain region-of-interest (ROI) imaging \cite{xia2013towards}, collected projection data may suffer from the data truncation problem \cite{huang2021data,wang2023review}. The primary cause of data truncation arises from two scenarios: a) The size of the imaged object exceeds the scanning field of view (FOV) due to the limited detector size \cite{huang2020field,khural2022deep,fonseca2021evaluation,Fan2022Fiducial,gao2023transformer};  b) In interior tomography, collimators are inserted to reduce unnecessary dose exposure for ROI imaging \cite{yu2006region,yu2008interior}.

Image reconstruction with the classic filtered back-projection (FBP) algorithm leads to cupping artifacts near the FOV boundary and missing (or severely distorted) anatomical structures outside the FOV boundary due to missing data \cite{huang2021data}. {Such degraded images limit practical applications in the clinical settings. For instance, patients need to undergo position verifications and radiotherapy planning base on their daily CT images for precise adaptive radiotherapy \cite{yang2025adaptive, zou2025predicting}. However, the sFOV images neither can identify the surface markers for treatment position verification or provide accurate electronic density for dose calculation \cite{gao2023transformer}.}
To solve this problem, the most straightforward idea is to recover the missing data in the projection domain. Following this idea, heuristic data interpolation emerged. This method can estimate the missing projection data outside the FOV based on the existing mathematical model or the geometric prior knowledge of the object. Therefore, it can alleviate the cupping artifacts near the FOV boundary by smoothing the transition between the measured data and the truncated data. Among them are representative methods such as symmetrical mirror interpolation \cite{ohnesorge2000efficient}, geometric shape-based data estimation \cite{hu2024application}, and water cylinder extrapolation (WCE) \cite{hsieh2004novel}.
With the rising of compressive sensing technology, iterative reconstruction algorithms with total variation regularization have been widely used to reconstruct images from insufficient projection data \cite{wang2023review}, such as sparse angle \cite{liu2014total} and limited angle imaging  \cite{huang2018scale} scenarios. For data truncation, minimizing the total variation loss can better reconstruct the images within the FOV \cite{yu2009compressed,han2009general}. However, all the above-mentioned FOV extension algorithms, including extrapolation methods and compressed sensing methods both, can only improve the image quality within the FOV, but cannot restore the missing anatomical structures outside the FOV effectively \cite{huang2021data}.

In the recent decades, deep learning algorithms have achieved astonishing results in CT imaging tasks \cite{huang2025learning, gao2024corediffu, xu2024prior}. For CT FOV extension, researchers have proposed various deep learning methods, which can generally be classified into image domain, projection domain, and dual-domain. a) Image domain: Khural et al. \cite{khural2022deep} applied the U-Net network to expand the visual field of the image after linear interpolation and improved the image quality. b) Projection domain: Fonseca \cite{fonseca2021evaluation} proposed a visual field expansion algorithm called HDeepFoV, which estimates the missing data outside the visual field through the FBPConvNet network and then combines the results with the measurement data in the projection domain to reconstruct the final image. Huang et al. \cite{huang2021data} proposed a Plug-and-Play framework, which improves the robustness and interpretability of the deep learning model by integrating measurement data consistency and the prior image information learned by the network. Tang et al. \cite{tang2024prior} adopted a similar idea and proposed Prior-FovNet, which uses the prior information of mega-voltage CT to expand the FOV of CBCT images. c) Dual-domain consistency: Gao et al. \cite{gao2023transformer} proposed a dual-domain visual field expansion algorithm based on transformer, which restores the missing anatomical structures outside the visual field by considering the consistency of the image domain and the projection domain. 


Diffusion models \cite{song2020score, ho2020denoising, dhariwal2021diffusion, nichol2021improved} are an emerging class of image generation models that degrade the image to a fully Gaussian noise by forward noise addition and obtain a high-quality target image conforming to the conditional distribution by sampling from known prior conditions through an inverse denoising process. The diffusion model has been widely used in image generation tasks due to its high image generation quality and stable training process, and has performed well for CT image reconstruction from insufficient data \cite{hou2024two,xie2024prior,zhang2024wavelet}. Nevertheless, the application of generative diffusion models to CT FOV extension has not been fully explored. Xie et al. \cite{xie2025score} proposed a null space shuttle algorithm based on score-based generative model, which decomposes a CT image with limited FOV into zero space and range space, and gradually recovers the missing anatomical structures in the zero space by using diffusion model guided by a prior information in the range space. Nevertheless, the FOV is extended by a new trajectory scan, whereas the diffusion model is applied to improve image quality inside the scan FOV. Liman et al. \cite{liman2024diffusion} proposed an image extrapolation method based on the conditional diffusion model Palette \cite{saharia2022palette}, which recovers missing anatomical structures outside the scan FOV for body composition analysis. These studies demonstrate the powerful generative capacity of diffusion models. 

Score-based generative models (SGMs) have demonstrated impressive sample quality and stable training performance in various image synthesis tasks \cite{yang2023diffusion, croitoru2023diffusion}. However, these models typically rely on a computationally intensive sampling process, requiring numerous iterative steps to progressively denoise Gaussian noise into realistic images. This slow sampling speed poses a major limitation for time-sensitive clinical applications \cite{zhou2024fast}. Recently, substantial efforts have been made to reduce the computational cost of training and inference in diffusion models. For instance, the conditional latent diffusion model (cLDM) \cite{rombach2022high,hou2024two} and the patch-based diffusion model (PatchDiffusion) \cite{wang2023patch} enable efficient training with large-size images. The diffusion generative adversarial network (DiffusionGAN) \cite{xiao2022DDGAN} enhances inference efficiency by adopting large sampling time steps using a multimodal Gaussian noise distribution.
An emerging alternative is diffusion modeling based on the Schr\"odinger bridge (SB) \cite{schrodinger1932theorie, de2021diffusion, shi2023diffusion, liu2023I2SB}, which enables direct transformation between two data distributions without the need to sample from a known noise prior. This framework naturally aligns with image-to-image translation tasks and offers a more efficient generative mechanism.
In this study, we explore the application of a Schr\"odinger bridge diffusion model, known as Image-to-Image Schr\"odinger Bridge (I$^2$SB) \cite{liu2023I2SB}, to the task of CT FOV extension. Unlike conventional diffusion methods that start from Gaussian noise, I$^2$SB directly learns the nonlinear stochastic process that maps limited-FOV image distributions to extended-FOV image distributions. This direct modeling leads to improved sampling efficiency and image fidelity. By reducing the number of required sampling steps while maintaining high generation quality, {I$^2$SB has potential to offer a clinically practical solution for CT FOV extension.}


\section{Materials and methods}

\subsection{Problem Formulation}

Generally speaking, all CT reconstruction tasks can be regarded as a process of solving an inverse problem. 
Here, we use $\boldsymbol{x} \in \mathbb{R}^m $ to represent the original image data and $\boldsymbol{y} \in \mathbb{R}^n $ to represent the measurement data. In an ideal scenario, the measurement data $\boldsymbol{y}$ can be expressed in terms of $\boldsymbol{x}$ using the following formula:
\begin{equation}
\boldsymbol{y} =\boldsymbol{A}\boldsymbol{x},
\label{eqn:FP}
\end{equation}
 where $\boldsymbol{A} \in \mathbb{R}^{m \times n}$ denotes the system matrix, which serves to project the image data $\boldsymbol{x}$ into the measurement data $\boldsymbol{y}$. 
 With data truncation, solving Eqn.\,(\ref{eqn:FP}) becomes a severely ill-posed inverse problem, and conventional analytic and iterative algorithms fail to reconstruct structures outside the scan FOV precisely.
 This implies that more prior knowledge about the scanned object is needed to assist the reconstruction process. Deep learning methods have more capacity in this regard using data-driven learning than conventional analytic or iterative reconstruction algorithms. A direct idea to solve the data truncation problem is using a deep learning model with learnable parameter $\boldsymbol{\theta}$ to establish the mapping from a limited FOV image $\boldsymbol{x}_s$ to a normal FOV (or FOV-extended) image $\boldsymbol{x}_l$, which can be expressed as: 
\begin{equation}
\label{deqn_exla}
\boldsymbol{x}_l = f({\boldsymbol{x}_s};\boldsymbol{\theta})
\end{equation}
Therefore, we can regard the task of restoring an image with limited FOV to an extended FOV image as an image-to-image generation task. In this work, we employ the I$^2$SB method \cite{liu2023I2SB} to accomplish this task. Moreover, the WCE reconstruction is used as $\boldsymbol{x}_s$ to reduce cupping artifacts and preliminarily restore some anatomical structures, which has been demonstrated to be more effective than a naive FBP reconstruction in our previous work \cite{huang2021data}. 

\subsection{Preliminaries in Diffusion Models}

SGMs \cite{yang2023diffusion, croitoru2023diffusion} are a class of powerful generative frameworks that model data distributions via stochastic differential equations (SDEs). Given a sample $\boldsymbol{x}_0$ from a known data distribution $p(\boldsymbol{x}_0)$, SGMs define a forward diffusion process that gradually corrupts the data with noise, and a reverse process that restores it.

In the forward process, noise is progressively added to a clean image $\boldsymbol{x}_0$ until it becomes pure Gaussian noise. This process is governed by the following SDE:
\begin{equation}
\label{eq:forward_sde}
\textrm{d}\boldsymbol{x}_t=\boldsymbol{f}_t(\boldsymbol{x}_t)\textrm{d}t+\sqrt{\beta_t}\textrm{d}\boldsymbol{w}_t,
\end{equation}
where $\boldsymbol{w}_t$ is a standard Wiener process, $\boldsymbol{f}_t(\boldsymbol{x}_t)$ is the drift coefficient, and $\beta_t$ is the diffusion coefficient, all potentially time-dependent. The time variable $t$ typically ranges from 0 to 1 and is discretized into $T$ steps. At $t=0$, $\boldsymbol{x}_0$ is the original data sample; at $t=1$, the corrupted sample $\boldsymbol{x}_1$ approximates a standard Gaussian distribution $\mathcal{N}(0, I)$.

To generate new data, SGMs define a reverse SDE that undoes this diffusion process:
\begin{equation}
\textrm{d}\boldsymbol{x}_t=[\boldsymbol{f}_t(\boldsymbol{x}_t)-\beta_t\nabla \log\,p(\boldsymbol{x}_t,t)]\textrm{d}t\,+\,\sqrt{\beta_t}\textrm{d}\bar{\boldsymbol{w}}_t
\label{eq:reverse_sde}
\end{equation}
where $p(\boldsymbol{x}_t, t)$ is the marginal distribution at time $t$, $\nabla \log p(\boldsymbol{x}_t, t)$ is its score function (i.e., the gradient of the log-probability), and $\bar{\boldsymbol{w}}_t$ is a standard Wiener process running backward in time. Given an initial Gaussian noise sample $\boldsymbol{x}_1$ and the score function at each time step, the clean image $\boldsymbol{x}_0$ can be recovered by integrating this reverse SDE.

To estimate the score function $\nabla \log p(\boldsymbol{x}_t, t)$, a denoising neural network $\epsilon_{\boldsymbol{\theta}}$ is trained using a scaled score matching objective:

\begin{equation}
\label{eq:score_matching}
||\epsilon(\boldsymbol{x}_t,t; \boldsymbol{\theta})-\sigma_t\nabla \log\,p(\boldsymbol{x}_t,t|\boldsymbol{x}_0)||,
\end{equation}
where $\sigma_t$ denotes the standard deviation of the conditional distribution $p(\boldsymbol{x}_t | \boldsymbol{x}_0)$, which is typically known or analytically tractable. Once training converges, the model can perform posterior sampling using the following recursive relation:
\begin{equation}
\label{eq:posterior_sampling}
\boldsymbol{x}_t \sim p(\boldsymbol{x}_t|\hat{\boldsymbol{x}}_0,\boldsymbol{x}_{t+1}),\;\; \boldsymbol{x}_{T} \sim \mathcal{N}(0, I),
\end{equation}
where the estimated clean image $\hat{\boldsymbol{x}}_0$ is computed from $\boldsymbol{x}_t$ and the network output as:
\begin{equation}
\hat{\boldsymbol{x}}_0 := \frac{\boldsymbol{x}_t - b_t \epsilon_{\boldsymbol{\theta}}(\boldsymbol{x}_t, t)}{a_t},
\end{equation}
based on the known formulation of the forward process, where $a_t$ and $b_t$ are process-specific coefficients. By recursively denoising through this reverse process, the model progressively reconstructs a clean image from Gaussian noise.

\subsection{Image-to-Image Schr\"odinger Bridge}
 
An SB is a probabilistic model that formulates optimal transport between two distributions under entropy regularization. {It finds the most probable (minimum-change) stochastic path between two endpoint distributions \cite{schrodinger1932theorie}.} It defines a pair of forward and backward SDEs that evolve between two boundary distributions. The dynamics of the SB are given by:
\begin{equation}
\label{deqn_exla}
\textrm{d}\boldsymbol{x}_t=\left[\boldsymbol{f}_t+\beta_t\nabla \log\Psi(\boldsymbol{x}_t, t)\right]\textrm{d}t + \sqrt{\beta_t} \textrm{d}\boldsymbol{w}_t,
\end{equation}
\begin{equation}
\label{deqn_exla}
\textrm{d}\boldsymbol{x}_t=\left[\boldsymbol{f}_t - \beta_t\nabla \log\widehat{\Psi}(\boldsymbol{x}_t, t)\right]\textrm{d}t + \sqrt{\beta_t}\textrm{d}\bar{\boldsymbol{w}}_t,
\end{equation}
where $\boldsymbol{x}_0$ represents the normal FOV CT image sampled from the target distribution $p_{l}$ in this work, while $\boldsymbol{x}_1$ represents the corresponding limited FOV CT image sampled from the source distribution $p_{s}$. The functions $\Psi$ and $\widehat{\Psi}$ represent time-dependent energy potentials that guide the transport process.
{When $\Psi(\boldsymbol{x}_t,t)$ and $\widehat{\Psi}(\boldsymbol{x}_t,t)$ are viewed as distribution densities, i.e., equivalent to $p(\boldsymbol{x}_t,t)$ in Eqn.\,(\ref{eq:reverse_sde}), $\nabla \log\widehat{\Psi}(x_t, t)$ and $\nabla \log \Psi(x_t, t)$ resemble the score functions of the following SDEs, respectively:}
\begin{equation}
\textrm{d}\boldsymbol{x}_t = \boldsymbol{f}_t(\boldsymbol{x}_t)\textrm{d}t+\sqrt{\beta_t}\textrm{d}\boldsymbol{w}_t, \boldsymbol{x}_0\sim \widehat{\Psi}(\cdot,0),
\end{equation}
\begin{equation}
\textrm{d}\boldsymbol{x}_t = \boldsymbol{f}_t(\boldsymbol{x}_t)\textrm{d}t+\sqrt{\beta_t}\textrm{d}\bar{\boldsymbol{w}}_t, \boldsymbol{x}_1\sim {\Psi}(\cdot,1).
\end{equation}

These potentials satisfy the following coupled partial differential equations (PDEs):
\begin{equation}
\begin{cases}
\dfrac{\partial \Psi(\boldsymbol{x}, t)}{\partial t} = -\nabla \Psi^{\top}\boldsymbol{f} - \dfrac{1}{2} \beta \nabla^2 \Psi ,\\
\dfrac{\partial \hat{\Psi}(\boldsymbol{x}, t)}{\partial t} = -\nabla \cdot (\hat{\Psi} \boldsymbol{f}) + \dfrac{1}{2} \beta \nabla^2 \hat{\Psi},
\end{cases}
\label{eqn:NLSDE}
\end{equation}
These two potentials $\Psi$ and $\widehat{\Psi}$ are connected through the Nelson's duality \cite{Nelson1967Dynamical}:
\begin{equation}
    \Psi(\boldsymbol{x}, t) \widehat{\Psi}(\boldsymbol{x}, t) = q(\boldsymbol{x}, t),
\end{equation}
where $q(\boldsymbol{x}, t)$ denotes the evolving intermediate density, equivalent to $p(\boldsymbol{x}, t)$ as used in the score-based diffusion formulation (see Eqn.~(\ref{eq:reverse_sde})).
As the SB framework seeks the most likely stochastic path connecting the two marginals $p_l$ and $p_s$, the following boundary constraints are imposed:
\begin{equation}
\Psi(\boldsymbol{x}, 0) \widehat{\Psi}(\boldsymbol{x}, 0) = p_{l}(\boldsymbol{x}), \Psi(\boldsymbol{x}, 1) \widehat{\Psi}(\boldsymbol{x}, 1) = p_{s}(\boldsymbol{x}).
\label{eq:constraints}
\end{equation}

Since Eqn.\,(\ref{eq:constraints}) establishes a two-way coupling between $\widehat{\Psi}(\cdot, 0)$ and $\Psi(\cdot, T)$, this means that when solving for $\widehat{\Psi}(\cdot, 0)$, $\Psi(\cdot, T)$ must also be taken into account simultaneously, complicating the problem-solving process. Therefore, according to \cite{liu2023I2SB}, $\widehat{\Psi}_0(\cdot)$ is assumed to be a Dirac $\delta$ distribution centered on each normal FOV CT image. With such an assumption, the initial distribution of the equation is given as:
\begin{equation}
\widehat{\Psi}(\cdot,0) =\delta_l(\cdot),\; \Psi(\cdot,1)=\frac{p_s}{\widehat{\Psi}(\cdot,1)}.
\end{equation}

\begin{figure}[!t]
\centering
\includegraphics[width=\linewidth]{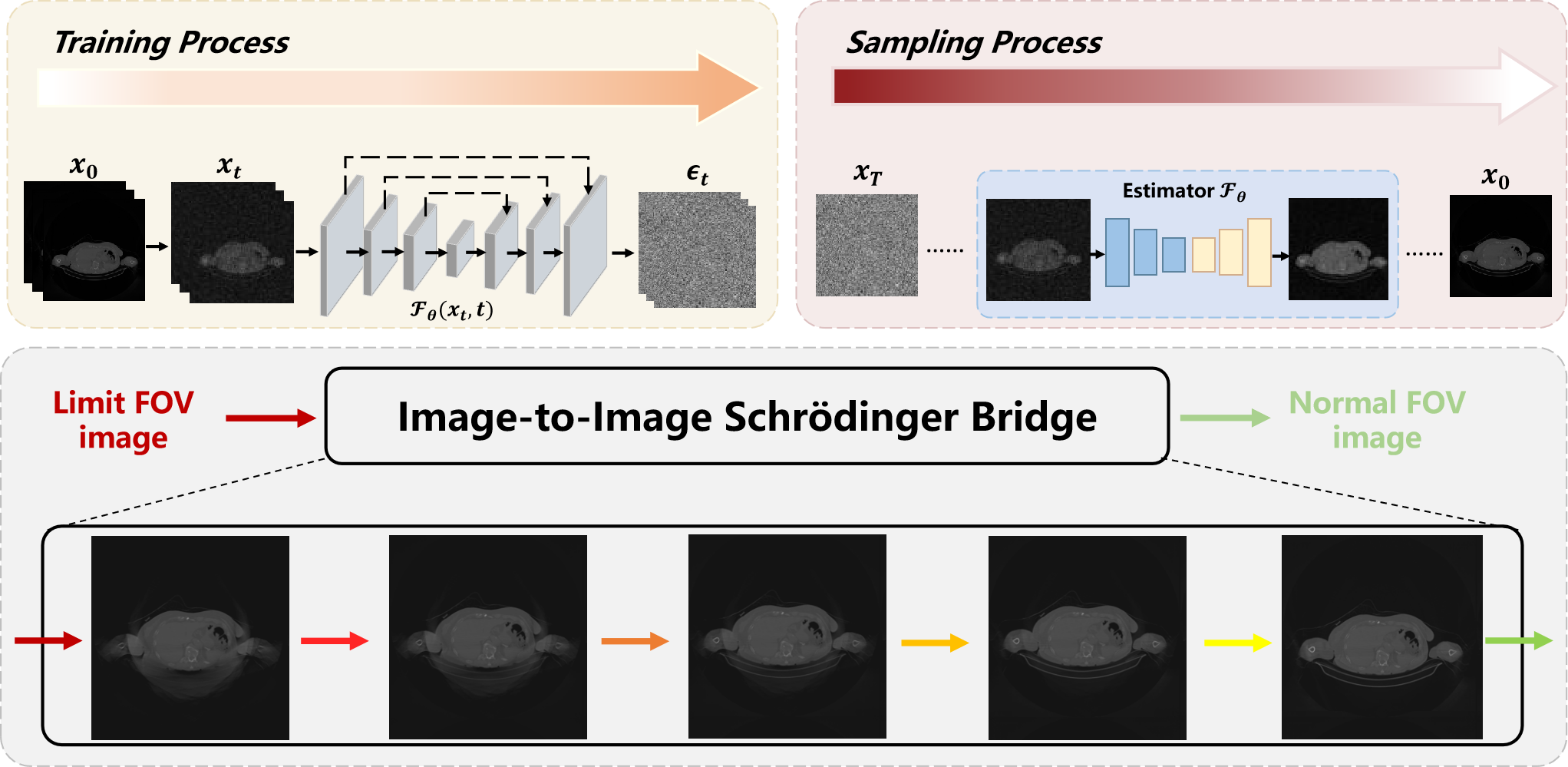}
\caption{The key difference in the image generation process between a conventional diffusion model and the image-to-image Schrödinger Bridge (I$^2$SB) model is that the former begins from a noise image, while I$^2$SB directly transforms an input image from one domain into its target counterpart. In this work, the limited FOV image is reconstructed by the water cylinder extrapolation (WCE) method \cite{huang2021data}.}
\label{fig_1}
\end{figure}

The above assumption breaks the dependency of $\widehat{\Psi}(\cdot, 0)$ on $\Psi(\cdot, T)$, which enables us to directly establish a diffusion bridge between the two distributions without having to fix the starting point of the diffusion as Gaussian noise. The differences between I$^2$SB and a conventional diffusion model architecture are shown in Fig.\,\ref{fig_1}. 

The specific process of applying the I$^2$SB model to the CT FOV extension task is introduced in the following. $\boldsymbol{x}_0$ represents an image with an extended FOV, while $\boldsymbol{x}_1$ represents the corresponding image with a limit FOV. Any $\boldsymbol{x}_t$ that drifts between $\boldsymbol{x}_0$ and $\boldsymbol{x}_1$ is designed to follow a Gaussian distribution $q(\boldsymbol{x}_t|\boldsymbol{x}_0, \boldsymbol{x}_1)$, which can be expressed as

\begin{equation}
q(\boldsymbol{x}_t|\boldsymbol{x}_0,\boldsymbol{x}_1)=\mathcal{N}(\boldsymbol{x}_t;\boldsymbol{\mu}_t({\boldsymbol{x}_0,\boldsymbol{x}_1}),\boldsymbol{\Sigma}_t),
\end{equation}
\begin{equation}
\boldsymbol{\mu}_t=\frac{\bar{\sigma}_t^2}{\bar{\sigma}_t^2+\sigma_t^2}\boldsymbol{x}_0 + \frac{\sigma_t^2}{\bar{\sigma}_t^2+\sigma_t^2}\boldsymbol{x}_1,\;\;
{\boldsymbol{\Sigma}_t}=\frac{\bar{\sigma}_t^2\sigma_t^2}{\bar{\sigma}_t^2+\sigma_t^2} \cdot \boldsymbol{I},
\end{equation}
where $\;\sigma_t^2:=\int_0^t\beta_\tau \textrm{d}\tau$ and $\bar{\sigma}_t^2:=\int_t^1\beta_\tau \textrm{d}\tau$ are the variances accumulated in the interval [0, 1] from either direction, respectively, and $\beta_\tau$ can determine the speed (or sampling step size) of the diffusion process. Since $\boldsymbol{x}_t$ can be obtained at any time-steps using the equation above, the denoising network can be trained by a simple loss function,
\begin{equation}
||\epsilon(\boldsymbol{x}_t,t;\boldsymbol{\theta})-\frac{\boldsymbol{x}_t - \boldsymbol{x}_0}{\sigma_t}||.
\end{equation}
Therefore, the mapping model $\boldsymbol{\theta}$ from a limit FOV CT image to a normal FOV image becomes a simple parameterized noise estimator $\mathcal{F}_\theta$.
Once the noise estimator can precisely predict noise level in any timestep, it can be used to help the iterative reconstruction process and the details are as follows. 

At any time step $k$ in the reverse process, e.g. from $\boldsymbol{x}_k$ to $\boldsymbol{x}_{k-1}$, $\hat{\boldsymbol{x}}_0$ can be calculated by the iterative last point $\boldsymbol{x}_k$ and the estimated noise $\epsilon_{\boldsymbol{\theta}}$,

\begin{equation}
\hat{\boldsymbol{x}}_0^k = \boldsymbol{x}_k - \sigma_{k} \epsilon_{\boldsymbol{\theta}}(\boldsymbol{x}_k,k).
\end{equation}

Once $\hat{\boldsymbol{x}}_0^k$ is acquired, $\boldsymbol{x}_{k-1}$ can be sampled from the posterior distribution $p(\boldsymbol{x}_{k-1}|\hat{\boldsymbol{x}}_0^k,\boldsymbol{x}_k)$, which can be expressed as:

\begin{equation}
p(\boldsymbol{x}_{k-1}|\hat{\boldsymbol{x}}_0^k,\boldsymbol{x}_k)=\mathcal{N}(\boldsymbol{x}_{k-1}; \frac{\alpha_{k-1}^2}{\sigma_k^2} \hat{\boldsymbol{x}}_0^k+\frac{\sigma_{k - 1}^2}{\sigma_k^2} \boldsymbol{x}_n, \frac{\sigma_{k - 1}^2 \alpha_{k - 1}^2}{\sigma_k^2} \boldsymbol{I}),
\end{equation}
where $\alpha_{k-1}^2=\int_{k-1}^{k}\beta_{\kappa}\textrm{d}\kappa$ implies the accumulated variance between the timestep $k-1$ to $k$. After repeating the sampling process $T$ times, we can obtain the final reconstructed image. However, the sampling process that involves $T$ sampling steps takes a long time to complete. We can adjust the number of steps for evaluating the number of neural functions (NFE) to balance the computational efficiency and the quality of the reconstructed image. 
The advantage of I$^2$SB is that NFE can be set to 1, which achieves high image generation efficiency without significant degradation of image quality.




\subsection{Experiment Setup}
\subsubsection{Simulated data}
The proposed I$^2$SB algorithm for CT FOV extension was evaluated on simulated noisy data. The SMIR dataset \cite{SMIRdata} containing head and neck CT data from 53 patients was used for these experiments. Since this is a public open access data repository, informed consent was not required for researchers to reuse this dataset for research purpose. A Siemens Artis zee angiographic C-arm system was simulated to generate CBCT projections. The system parameters are displayed in Table \ref{tab:simulate_params}. Poisson noise was simulated assuming an incident photon number of $10^5$ (i.e., $I_0= 10^5$) for each detector pixel. The 3D Feldkamp-Davis-Kress (FDK) reconstruction with WCE \cite{hsieh2004novel} was used to compute the reconstruction from truncated projection data. 
The reconstructed volumes have a size of 512 $\times$ 512 $\times$ 512 with a voxel size of 1.27 mm $\times$ 1.27 mm $\times$ 1.27 mm. 
The original scan FOV diameter is 34 cm. The extended FOV images have a diameter of 65 cm. {To improve data representativeness while avoiding redundancy from highly similar adjacent slices, each CT sequence was downsampled by selecting one slice every 10 axial slices. Importantly, the training, validation, and test sets are composed of slices from distinct patients to ensure the independence and reliability of the evaluation.} 2050 slices from 41 patients were used for training, 50 slices from one patient was used for validation, and 500 slices from 10 patients were used for testing. As a proof-of-concept study, 512 $\times$ 512 images were rebinned to 256 $\times$ 256 images to save computational resources and experimental time. 

\begin{table}[t]
\centering
\caption{Parameters for simulating imaging system in experiments.}
\label{tab:simulate_params}
\begin{tabular}{cc}
\toprule
Parameters & Values \\
\midrule
Source-to-detector distance $d_1$ (mm) & 900 \\
Source-to-isocenter distance $d_2$ (mm) & 600 \\
Size of detector  (pixels) & 1240 $\times$ 960 \\
Pixels size (mm) & 1.0 \\
FOV diameter (cm) & 34 \\
Reconstruction volume size (voxels) & 512$\times$512$\times$512 \\
Voxel size (mm) & 1.27$\times$1.27$\times$1.27 \\
\bottomrule
\end{tabular}
\end{table}

\subsubsection{Real data}
To better verify the usability of the proposed algorithm, an evaluation on {real head data of one patient} with a typical noise level was conducted as well. The head {data} was collected using the Artis zee vascular angiography C-arm system (Siemens Healthcare GmbH, Forchheim). Since this data has been scanned for prior projects \cite{huang2021data}, ethical review and approval was not required for this study in accordance with the local legislation and institutional requirements (BayKrG Art. 27). The dose area product (DAP) of the entire scan was 532 $\mu Gy\cdot m^2$. The detector size was 1240 $\times$ 960 pixels, and the detector pixel size was 0.308\,mm$\,\times\,$0.308\,mm. The entire dataset contained 496 {projection views} obtained from a 200$^\circ$ short scan. The reconstructed volume size was 512\,$\times$\,512\,$\times$ 300 with an isotropic voxel size of 0.465\,mm$\,\times\,$0.465\,mm$\,\times\,$0.465\,mm. {The reference volume was reconstructed using the fast iterative shrinkage-thresholding algorithm (FISTA) with iterative reweighted total variation regularization \cite{huang2018scale} from this non-truncated projection data.}

{For the ROI imaging setup, we simulated a virtual collimator by using only the central 600\,$\times$\,960 pixels of the original projection data. All pixels on both sides of this central region were set to zero, effectively reducing the reconstruction FOV diameter from 23.9\,cm to 11.6\,cm. All the 300 axial slices from the WCE reconstruction were used for test.}

In this study, the model was trained from simulated data and evaluated on the above real head data. In total 491 patients from the CQ500 head CT data collection \cite{CHILAMKURTHY20182388} were used to simulate WCE reconstruction and full reconstruction image pairs {from generated projection data} in the same C-arm system as the real data to train the model. Among these 3D reconstructions, 2760 slices from 471 patients were used for training, and 120 slices from the other 20 patients were used for validation. {Since the projection data was generated from the CQ500 head CT volumes, to} achieve a similar noise level, Poisson noise was added to the projection data, considering the exposure before attenuation for each detector pixel to be $I_0=5\times10^5$ photons. 

\begin{figure*}[b]
\centering
{\includegraphics[width=0.16\textwidth]{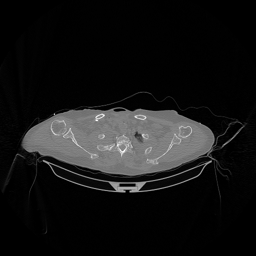}\label{fig_n1_gt}}
\hfil
{\includegraphics[width=0.16\textwidth]{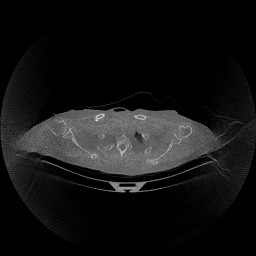}\label{fig_n1_wce}}
\hfil
{\includegraphics[width=0.16\textwidth]{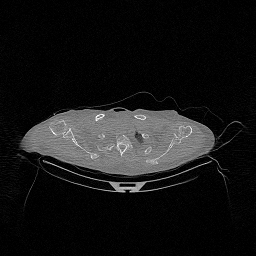}\label{fig_n1_ddpm}}
\hfil
{\includegraphics[width=0.16\textwidth]{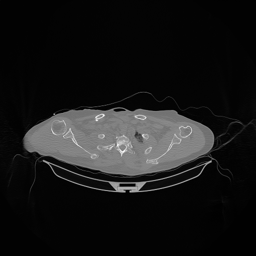}\label{fig_n1_patchdiffu}}
\hfil
{\includegraphics[width=0.16\textwidth]{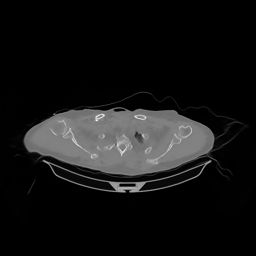}\label{fig_n1_diffGAN}}
\hfil
{\includegraphics[width=0.16\textwidth]{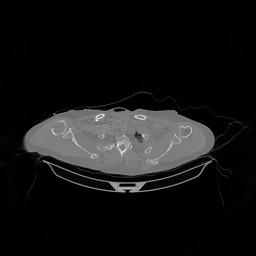}}\label{fig_n1_proposed}
\hfil

{\includegraphics[width=0.16\textwidth]{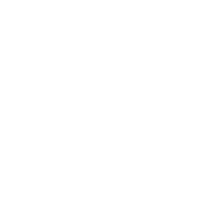}\label{fig_n1_gt_e}}
\hfil
{\includegraphics[width=0.16\textwidth]{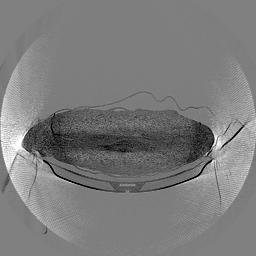}\label{fig_n1_wce_e}}
\hfil
{\includegraphics[width=0.16\textwidth]{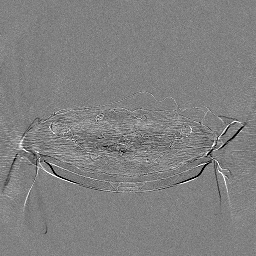}\label{fig_n1_ddpm_e}}
\hfil
{\includegraphics[width=0.16\textwidth]{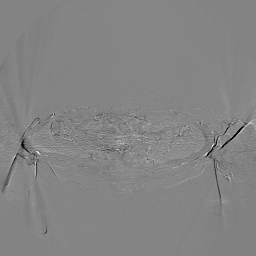}\label{fig_n1_patchdiffu_e}}
\hfil
{\includegraphics[width=0.16\textwidth]{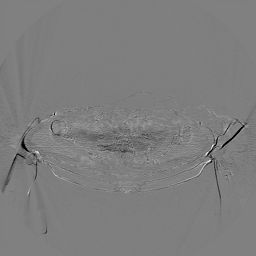}\label{fig_n1_diffGAN_e}}
\hfil
{\includegraphics[width=0.16\textwidth]{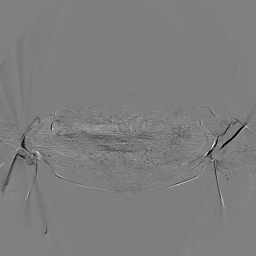}\label{fig_n1_proposed_e}}
\hfil

{\includegraphics[width=0.16\textwidth]{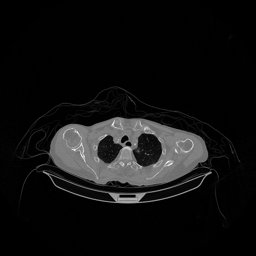}\label{fig_n2_gt}}
\hfil
{\includegraphics[width=0.16\textwidth]{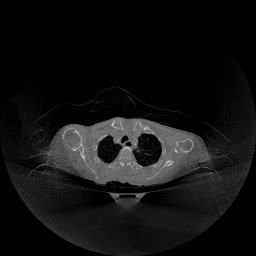}\label{fig_n2_wce}}
\hfil
{\includegraphics[width=0.16\textwidth]{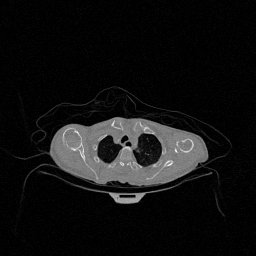}\label{fig_n2_ddpm}}
\hfil
{\includegraphics[width=0.16\textwidth]{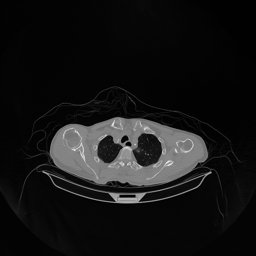}\label{fig_n2_patchdiffu}}
\hfil
{\includegraphics[width=0.16\textwidth]{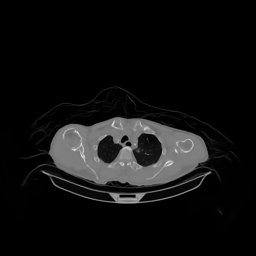}}
\hfil
{\includegraphics[width=0.16\textwidth]{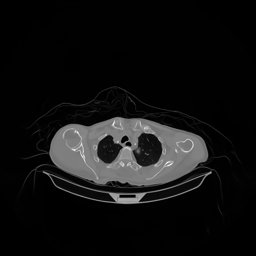}\label{fig_n2_proposed}}
\hfil

\subfloat[GT]{\includegraphics[width=0.16\textwidth]{blank.png}\label{fig_n2_gt_error}}
\hfil
\subfloat[WCE (input)]{\includegraphics[width=0.16\textwidth]{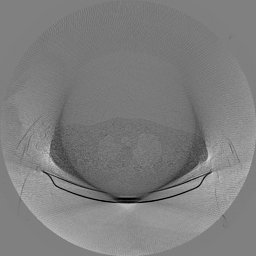}\label{fig_n2_wce_error}}
\hfil
\subfloat[cDDPM]{\includegraphics[width=0.16\textwidth]{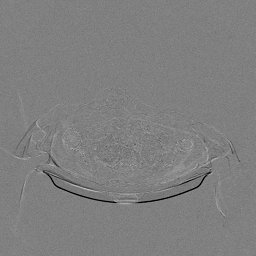}\label{fig_n2_ddpm_error}}
\hfil
\subfloat[PatchDiffusion]{\includegraphics[width=0.16\textwidth]{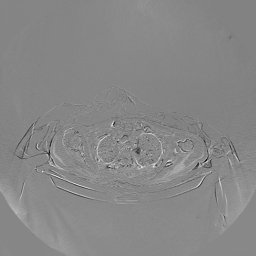}\label{fig_n2_patchdiffu_error}}
\hfil
\subfloat[DiffusionGAN]{\includegraphics[width=0.16\textwidth]{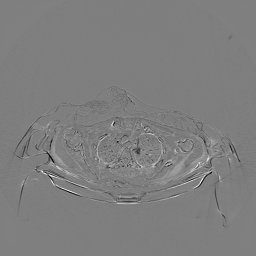}\label{fig_nw_diffGAN_error}}
\hfil
\subfloat[\textbf{I$^2$SB}]{\includegraphics[width=0.16\textwidth]{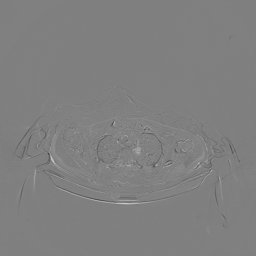}\label{fig_n2_proposed_error}}
\hfil

\caption{Results of two exemplary test slices in the noisy scenario. The first and third rows represent different slices in the test set, and the second and fourth rows show the error maps. The display windows are [-1000, 1000]\,HU for the reconstruction images and [-500, 500]\,HU for the error maps, respectively.}
\label{fig_noise_comparison}
\end{figure*}

\subsubsection{Network parameters}
All the experiments were conducted on a GPU server with GeForce A6000 GPUs. The WCE images were used as the input images of the learning-based methods, which have advantages over FBP images as the input as found in our previous study \cite{huang2021data}. 
For the proposed I$^2$SB algorithm, the ordinary differential equation (ODE) mode was applied. A symmetric scheduling $\beta_t$ was set with the total time-steps $K=1000$. The network was initialized with the unconditional ADM checkpoint \cite{dhariwal2021diffusion} trained on ImageNet $256\times256$. 
Followed by \cite{liu2023I2SB}, the maximum of hyperparameter $\beta_t$ was set to 0.3. The model was trained and tested at a size of 256 × 256 images. A batch size of 32 and a micro batch size of 4 were employed during training, using the Adam optimizer with a learning rate of $5\times10^{-5}$ for 200,000 iterations.

\subsubsection{Comparison models}
To verify the performance of the proposed I$^2$SB algorithm, we compared it with several prominent image-to-image translation algorithms such as the FBPConvNet \cite{jin2017deep}, Pix2pixGAN \cite{isola2017image}, the conditional denoising diffusion probabilistic model (cDDPM) \cite{peng2024cbct},  PatchDiffusion \cite{wang2023patch}, and DiffusionGAN \cite{xiao2022DDGAN}. To ensure network convergence, the number of training epochs for FBPConvNet and Pix2pixGAN was set to 200. While cDDPM and PatchDiffusion are based on the original DDPM architecture, the condition images in these models were added by concatenate on channel dimension, and their training iterations were set to $1\times10^6$. {The patch size of PatchDiffusion was $64 \time 64$ with a stride of 32 pixels.}

This article used three common objective image evaluation parameters, including root mean square error (RMSE), peak signal-to-noise ratio (PSNR), and structural similarity index (SSIM), to evaluate image quality. {For the SSIM computations, all CT images were windowed to a range of [-1000, 1000] HU and then converted to 8-bit images. After this preprocessing, a default window size of 7 pixels and a dynamic range of 255 were applied.}

\section{Results}

\subsection{Results of Simulated Noisy Data}

The results for simulated noisy data are shown in Fig.\,\ref{fig_noise_comparison}. The first and third rows display two representative slices from the test set, and the second and fourth rows present the corresponding error maps. The traditional WCE method suppresses cupping artifacts and recovers some missing anatomical structures but still shows noticeable deviations from the ground truth. 
Predictions from the four diffusion-based models demonstrate a markedly improved ability to restore anatomical structures. Among them, cDDPM fails to fully reconstruct the patient bed and retains residual noise in its outputs. This noise is attributable to an incomplete reverse denoising process rather than residual Poisson noise, as evidenced in our noise-free experiments (Fig.\,\ref{fig_noise_free_comparison} in the Appendix). {PatchDiffusion, DiffusionGAN, and I$^2$SB achieve similar visual quality in both the anatomical region and the patient bed. In the background region, PatchDiffusion is capable of reducing noise and artifacts to a similar extent as DiffusionGAN and I$^2$SB, as illustrated by the top row difference images. Nevertheless, a major portion of PatchDiffusion predictions still exhibit noticeable noise and residual artifacts in the background area with a narrow window of [-500, 500]\,HU for the error maps, as shown in the bottom difference image of Fig.\,\ref{fig_noise_comparison}(d).}

Quantitative results in {Tab.\,\ref{tab:metrics_n}} further confirm I$^2$SB’s superiority over conventional deep learning methods such as FBPConvNet and Pix2pixGAN across RMSE, PSNR, and SSIM (p $<$ 0.01 for RMSE, PSNR and SSIM using \modified{slice-wise paired} t-test). Compared to other diffusion models such as cDDPM and PatchDiffusion (p $<$ 0.01), I$^2$SB delivers higher image quality. {It is worth noting that PatchDiffusion achieves reasonable RMSE and PSNR values, but got a very low mean SSIM value of 0.5636, potentially caused by the low-magnitude artifacts and noise in the background. DiffusionGAN achieves comparable quantitative performance to I$^2$SB without statistical significance for PSNR (p = 0.075 $>$ 0.05 for PSNR, p $<$ 0.01 for RMSE and SSIM). However, I$^2$SB demonstrates a significant advantage in inference efficiency, as summarized in Tab.\,\ref{tab:infer_time}.

\begin{table}[ht]
  \centering
  \caption{{Quantitative comparison in the noisy scenario.}}
  \begin{tabular}{lcccc}
    \toprule
    \text{Methods} & RMSE & PSNR $\;$ & {SSIM}  \\
    \midrule
    WCE  & 168.4 $\pm$ 33.5 & 22.29 $\pm$ 1.86 & 0.5235  $\pm$ 0.079\\
    Pix2pixGAN & 110.8 $\pm$ 15.1 & 24.76 $\pm$ 1.91 & 0.7146 $\pm$ 0.071 \\
    FBPConvNet  & 60.8 $\pm$ 14.2 & 28.79 $\pm$ 2.00 & 0.8433 $\pm$ 0.061\\
    cDDPM  & 76.8 $\pm$ 16.4 & 28.78 $\pm$ 1.78 & 0.6799  $\pm$ 0.027\\   
    PatchDiffusion & 65.2 $\pm$ 15.0 & 30.11 $\pm$ 1.98 & 0.5636 $\pm$ 0.059\\
    DiffusionGAN & 48.0 $\pm$ 17.0 & 32.89 $\pm$ 2.69 & 0.9079 $\pm$  0.048\\
    I$^2$SB (proposed)  & 49.8 $\pm$ 18.8 & 32.73 $\pm$ 3.21 & 0.9021 $\pm$  0.053\\
    \bottomrule
  \end{tabular}
  \label{tab:metrics_n} 
\end{table}

\begin{figure*}[b]
\centering
\includegraphics[width=0.65\linewidth]{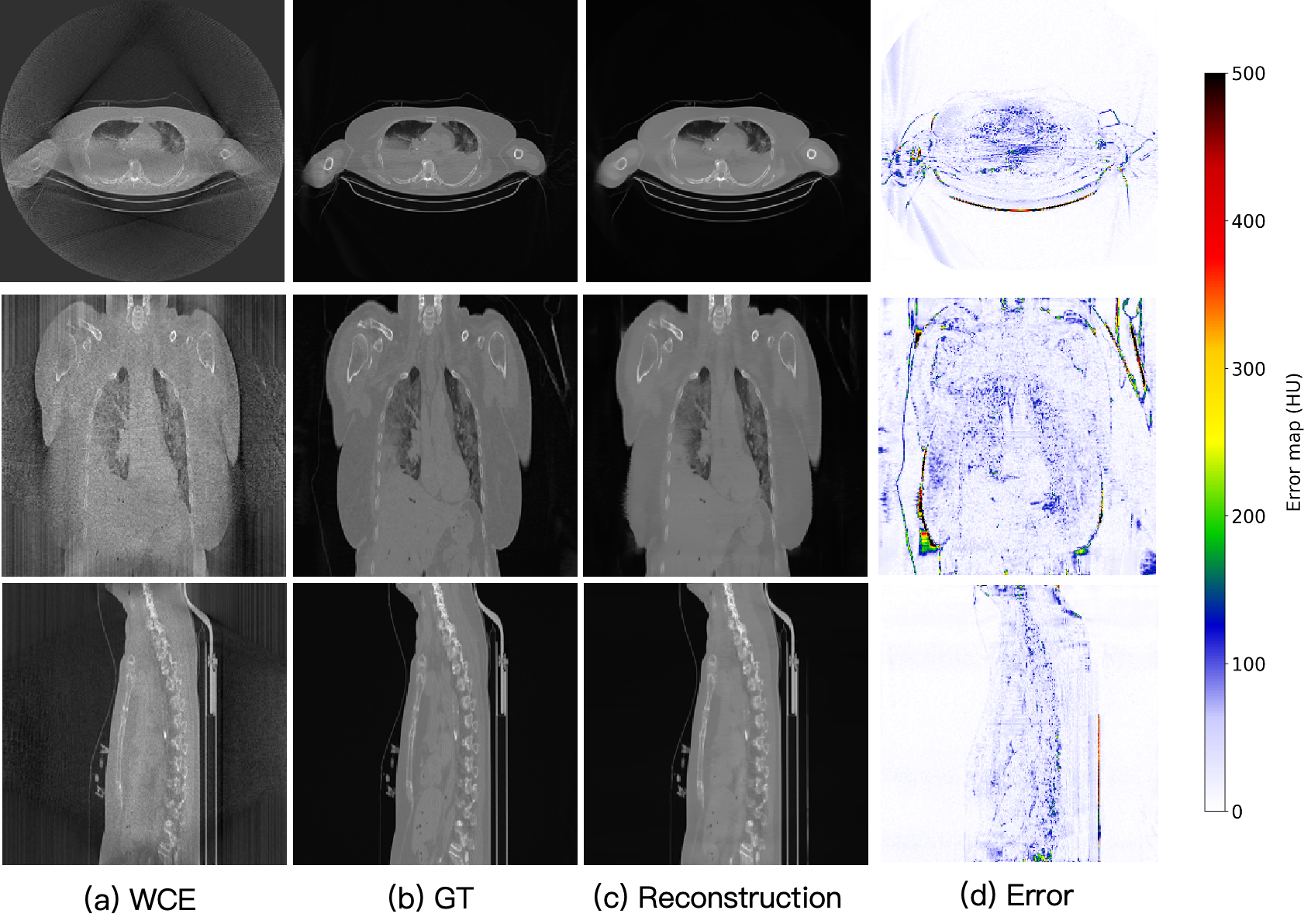}
\caption{{The FOV extension images of one case from axial, coronal and sagittal views. Central slices were selected as the representative slices: (a) WCE, (b) ground truth (GT), (c) I$^2$SB, (d) the difference between I$^2$SB reconstruction and GT. The display ranges are [-1000, 1000]\,HU for the reconstructions and [0, 500] HU for the absolute errors, respectively.}}
\label{fig_pred_and_gt}
\end{figure*}

\subsection{Efficiency Analysis}

\begin{table}[h]
\centering
\caption{{The influence of NFE on image quality.}}
\label{tab:nfe_eval}
\begin{tabular}{lccc}
\toprule
NFE & RMSE & PSNR & SSIM \\
\midrule
1   & 49.8 $\pm$ 18.8 & 32.73 $\pm$ 3.20& 0.9021 $\pm$ 0.053  \\
5   & 50.6 $\pm$ 19.4 & 32.64 $\pm$ 3.28 & 0.9031 $\pm$ 0.051 \\
10  & 50.4 $\pm$ 19.4 & 32.67 $\pm$ 3.27& 0.9029 $\pm$ 0.050 \\
25  & 50.6 $\pm$ 19.4 & 32.64 $\pm$ 3.28& 0.9014 $\pm$ 0.051 \\
\bottomrule
\end{tabular}
\end{table}

\begin{table}[h]
\centering
\caption{{Inference time comparison using different methods}}
\begin{tabular}{lc}
\toprule
\text{Method} & Time (s) \\
\midrule
FBPConvNet & 0.22 \\
Pix2pixGAN & 0.43 \\
cDDPM & 135.46 \\
PatchDiffusion & 6.24 \\
DiffusionGAN & 0.57 \\
\textbf{I$^2$SB} & \textbf{0.19} \\
\bottomrule
\end{tabular}
\label{tab:infer_time}
\end{table}


We further conducted reconstructions using the model with different NFEs. Here, in order to enhance the assessment efficiency, the number of test images was set to 250. The statistical results of the evaluation metrics are presented in Tab. \ref{tab:nfe_eval}.
When NFE increases, image quality has minor improvement without statistical significance. However, as NFE increases, the sampling time for each image rapidly rises, reducing the reconstruction efficiency. 
Therefore, one-step generation can be regarded as the optimal number of NFEs in this task.

The inference times per 2D slice for different methods are summarized in Tab.\,\ref{tab:infer_time}. FBPConvNet and Pix2pixGAN, which require no iterative sampling, achieve inference times below 0.5\,s. In contrast, the cDDPM model takes over 2 minutes due to its reliance on hundreds of reverse sampling steps. The PatchDiffusion model shortens this to 6.2\,s, while DiffusionGAN, with only four reverse sampling steps, achieves 0.57\,s. Benefiting from a single-step sampling process, I$^2$SB attains the fastest inference time among all diffusion-based methods, at just 0.19\,s.

\begin{figure*}[b]
\centering
\includegraphics[width=\linewidth]{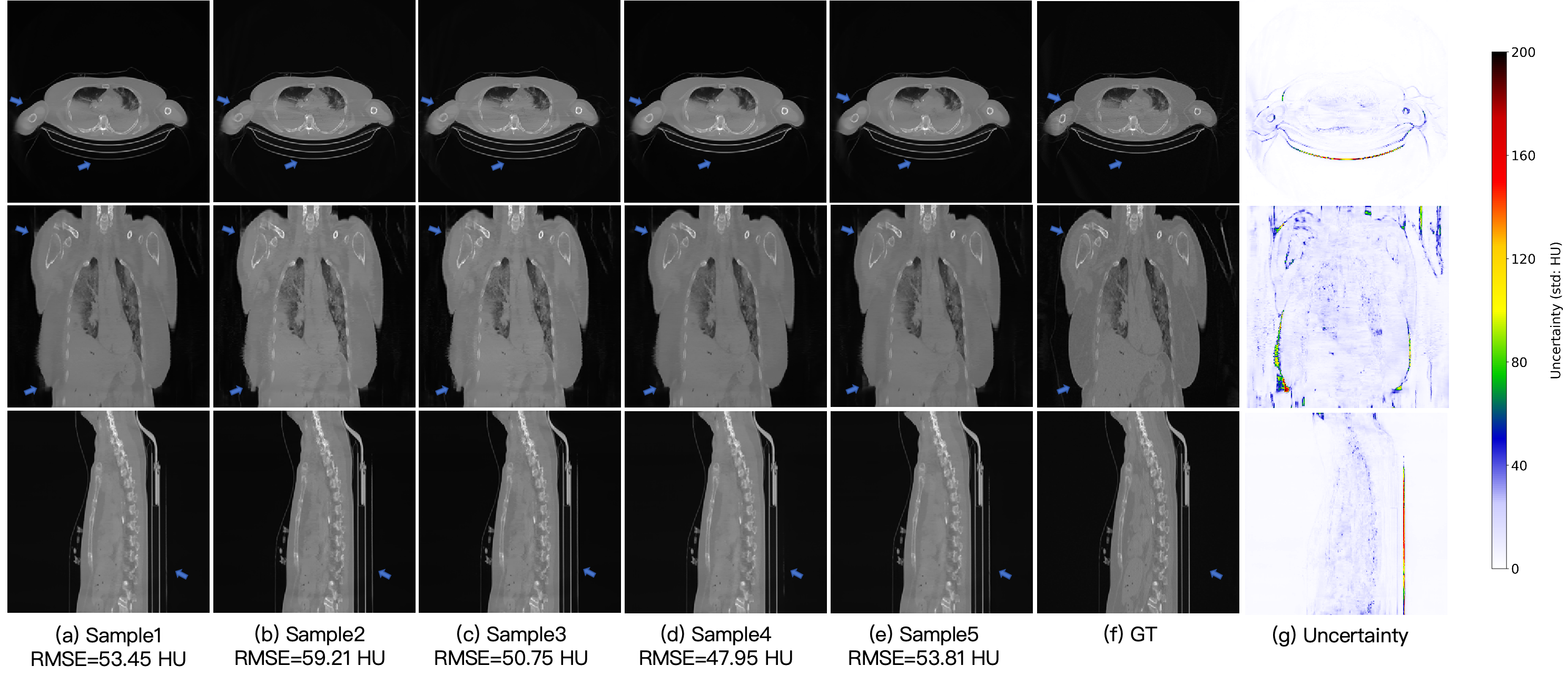}
\caption{{Quantifying the uncertainty of using different seeds for training. The same case was used for uncertainty analysis. (a-e) training models with different seeds and sampling images with the same random seeds, (f) ground truth, (g) the standard deviation of the samples. The blue arrows point out reconstruction variances in the samples. The display windows are [-1000, 1000]\,HU for the reconstruction images and [0, 200]\,HU for the uncertainty maps, respectively.}}
\label{fig_uncertainty_different_seeds}
\end{figure*}

\subsection{Consistency and Uncertainty Analysis}


The proposed method is trained with two-dimensional (2D) axial slices, while the truncation problem is a three-dimensional (3D) issue. When using a 2D approach to attempt to solve a 3D problem, context discontinuity problems. Therefore, in order to evaluate the inter-layer continuity of the proposed method, we present the axial, coronal,  sagittal, and the difference images of one case, as shown in Fig.\,\ref{fig_pred_and_gt}. We can observe that some blurry details occur in some tissues at the FOV boundary, and our model also has error in reconstructing the bed plate, which are drawbacks of I$^2$SB. Nevertheless, I$^2$SB not only has realistic reconstruction appearance in the axial plane, but also maintains decent inter-layer continuity in other intersectional planes, as demonstrated by Fig.\,\ref{fig_pred_and_gt}. Extending the 2D I$^2$SB to 2.5D or 3D versions can potentially improve the inter-layer continuity at the cost of increased computational complexity. 

The classical diffusion model reconstructs images through the progressive removal of initially applied Gaussian noise. This inherently stochastic process introduces uncertainty, owing to its reliance on randomly sampled noise for initialization. In contrast, I$^2$SB directly learns the diffusion bridge between two sample distributions, thereby bypassing the need for initial random noise sampling and reducing variability in reconstruction. To assess this reduction, we first trained the model under a fixed random seed and performed multiple reconstructions with varied seeds. The results exhibited only negligible differences. To evaluate uncertainty more comprehensively, we then trained separate models under different random seeds and performed inference using the same seeds. As shown in Fig.\,\ref{fig_uncertainty_different_seeds}, the extended FOV reconstruction maintains a high degree of consistency for the same image slice. Training with different seeds leads to slight variations in the model’s recovery of anatomical structures outside the original FOV. Nonetheless, the RMSE across all seeds remains between 50\,HU and 60\,HU, indicating a similar overall error level.


\subsection{Real Clinical Data Experiment}
The experimental results on clinical head data are shown in Fig.\,\ref{fig_real_data}. The reference images were reconstructed using FISTA with iterative reweighted total variation regularization from non-truncated projection data. In the WCE reconstructions (Fig.\,\ref{fig_real_data}(b)), severe truncation prevents accurate recovery of anatomical structures outside the FOV. Despite being trained solely on simulated data with a domain gap, all deep learning models can restore a substantial portion of the missing anatomy. According to the quantitative analysis in Tab.\,\ref{tab:metrics_real}, the diffusion-based methods generally achieve better RMSE, PSNR, and SSIM metrics than the conventional deep learning approaches such as FBPConvNet and Pix2pixGAN, highlighting their stronger image generation capability.
However, cDDPM reconstructions exhibit more noticeable noise than those from other methods, consistent with the simulated data results. The PatchDiffusion model introduces artifacts within the FOV, likely due to its patch-wise processing strategy. While I$^2$SB and DiffusionGAN share the same limitations as other diffusion models in perfectly restoring soft-tissue detail, they produce fewer residual noise patterns and fewer artifacts within the FOV boundaries. The quantitative results in Tab.\,\ref{tab:metrics_real} shows that I$^2$SB achieves slightly better image quality than DiffusionGAN, in addition to the faster inference time. Overall, Fig.\,\ref{fig_real_data} demonstrates the strong efficacy of I$^2$SB in reconstructing real CBCT data.

\begin{figure*}[!t]
\centering
{\includegraphics[width=0.16\textwidth]{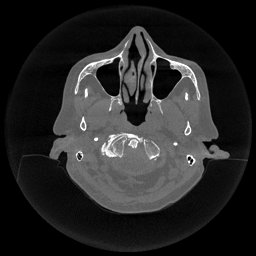}\label{fig_r1_gt}}
\hfil
{\includegraphics[width=0.16\textwidth]{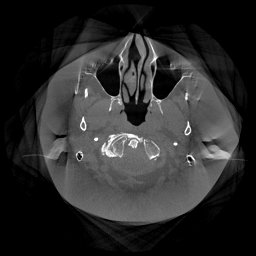}\label{fig_r1_wce}}
\hfil
{\includegraphics[width=0.16\textwidth]{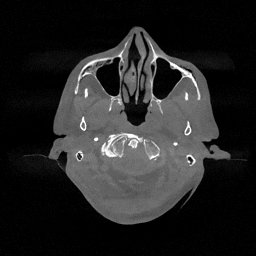}\label{fig_r1_ddpm}}
\hfil
{\includegraphics[width=0.16\textwidth]{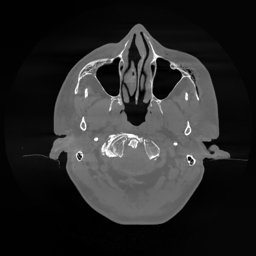}\label{fig_n1_patchdiffu}}
\hfil
{\includegraphics[width=0.16\textwidth]{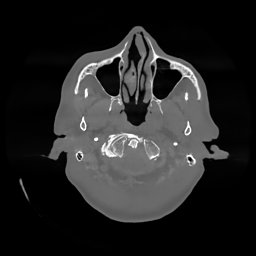}\label{fig_n1_DiffusionGAN}}
\hfil
{\includegraphics[width=0.16\textwidth]{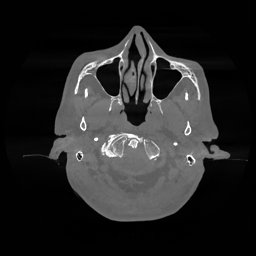}}\label{fig_r1_proposed}
\hfil

{\includegraphics[width=0.16\textwidth]{blank.png}\label{fig_r1_gt_e}}
\hfil
{\includegraphics[width=0.16\textwidth]{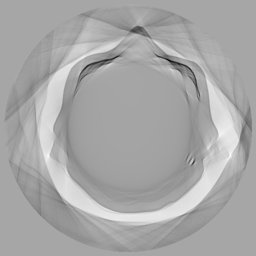}\label{fig_r1_wce_e}}
\hfil
{\includegraphics[width=0.16\textwidth]{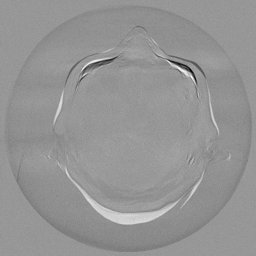}\label{fig_r1_ddpm_e}}
\hfil
{\includegraphics[width=0.16\textwidth]{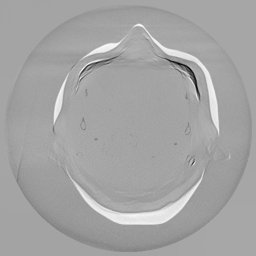}\label{fig_n1_patchdiffu_e}}
\hfil
{\includegraphics[width=0.16\textwidth]{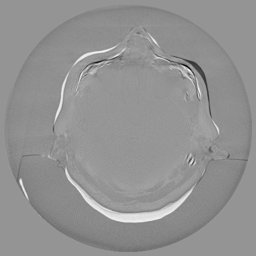}\label{fig_n1_DiffusionGAN_e}}
\hfil
{\includegraphics[width=0.16\textwidth]{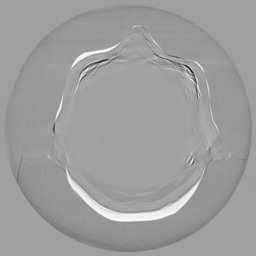}\label{fig_r1_proposed_e}}
\hfil

{\includegraphics[width=0.16\textwidth]{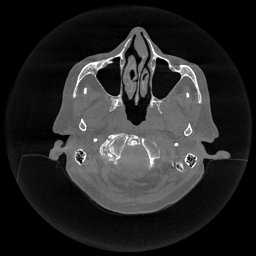}\label{fig_r2_gt}}
\hfil
{\includegraphics[width=0.16\textwidth]{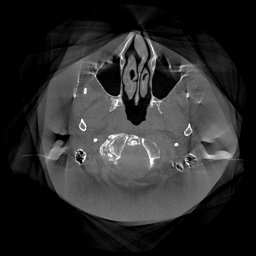}\label{fig_r2_wce}}
\hfil
{\includegraphics[width=0.16\textwidth]{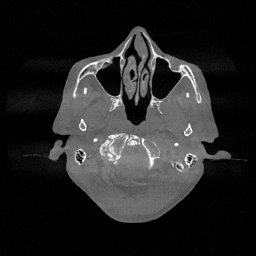}\label{fig_r2_ddpm}}
\hfil
{\includegraphics[width=0.16\textwidth]{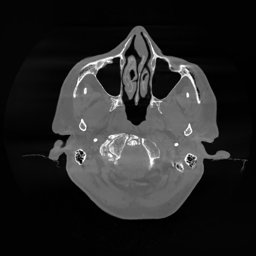}\label{fig_r2_patchdiffu}}
\hfil
{\includegraphics[width=0.16\textwidth]{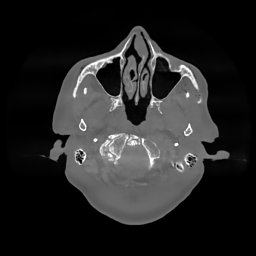}\label{fig_r2_DiffusionGAN}}
\hfil
{\includegraphics[width=0.16\textwidth]{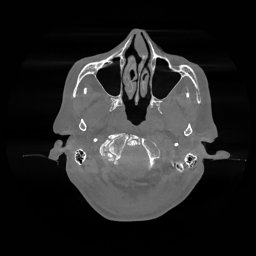}\label{fig_r2_proposed}}
\hfil

\subfloat[{Reference}]{\includegraphics[width=0.16\textwidth]{blank.png}\label{fig_r2_gt_error}}
\hfil
\subfloat[WCE (input)]{\includegraphics[width=0.16\textwidth]{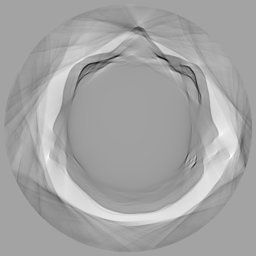}\label{fig_r2_wce_error}}
\hfil
\subfloat[cDDPM]{\includegraphics[width=0.16\textwidth]{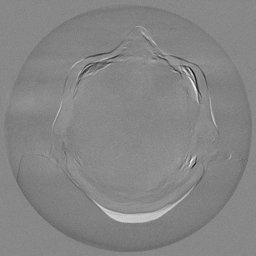}\label{fig_r2_ddpm_error}}
\hfil
\subfloat[PatchDiffusion]{\includegraphics[width=0.16\textwidth]{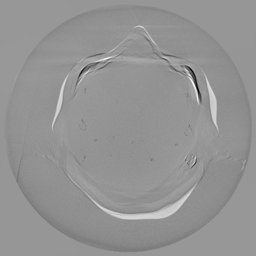}\label{fig_n2_patchdiffu_error}}
\hfil
\subfloat[DiffusionGAN]
{\includegraphics[width=0.16\textwidth]{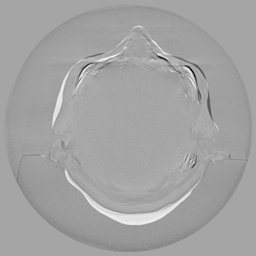}\label{fig_n2_DiffusionGAN_error}}
\hfil
\subfloat[\textbf{I$^2$SB}]{\includegraphics[width=0.16\textwidth]{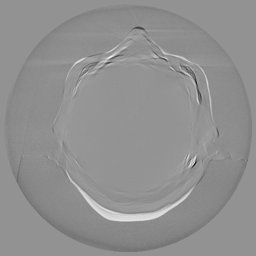}\label{fig_r2_proposed_error}}
\hfil

\caption{Results of two exemplary test slices in the real data. The first and third rows represent different slices in the test set, and the second and fourth rows show the error maps. The display window is [-1000, 1000]\,HU for the reconstruction images and the error maps.}
\label{fig_real_data}
\end{figure*}

\begin{table}[ht]
  \centering
  \caption{{Quantitative comparison in the {real data}.}}
  \begin{tabular}{lcccc}
    \toprule
    \text{Methods} & RMSE & PSNR $\;$ & {SSIM}  \\
    \midrule
    WCE  & 382.2 $\pm$ 102.2 & 15.40 $\pm$ 3.43 & 0.4221 $\pm$ 0.049  \\
    Pix2pixGAN & 279.8 $\pm$ 43.2 & 18.17 $\pm$ 1.30 & 0.4761 $\pm$ 0.020 \\
    FBPConvNet  & 260.0 $\pm$ 62.6 & 18.55 $\pm$ 2.27 & 0.5413 $\pm$ 0.049 \\
    cDDPM  & 190.2 $\pm$ 33.2 & 22.07 $\pm$ 2.04 & 0.5750 $\pm$ 0.039 \\
    PatchDiffusion & 231.8 $\pm$ 63.2 & 19.84 $\pm$ 2.86 & 0.5109 $\pm$ 0.070 \\
    DiffusionGAN  &   162.1 $\pm$ 25.4 & 22.78 $\pm$ 1.55 & 0.5884 $\pm$ 0.037 \\
    I$^2$SB (proposed)  & 152.0 $\pm$ 35.2 & 23.18 $\pm$ 2.47 & 0.6400 $\pm$ 0.045 \\
    \bottomrule
  \end{tabular}
  \label{tab:metrics_real} 
\end{table}

\section{Discussion}
In this work, we investigated several diffusion models for the task of CT FOV extension, formulated as an image-to-image translation problem. Conventional diffusion models have demonstrated strong performance in image generation, but their practical application is constrained by long inference times and high training costs. The proposed I$^2$SB model addresses these limitations by directly establishing a diffusion bridge between the limited-FOV and extended-FOV images. Unlike classical diffusion models, I$^2$SB does not require transforming the image into pure Gaussian noise; instead, inference begins from the truncated image itself. This approach reduces the number of sampling steps required while better preserving anatomical structure consistency. Our results show that, even with a single sampling step, I$^2$SB outperforms traditional reconstruction methods such as Pix2pixGAN and FBPConvNet, as well as other diffusion models including cDDPM and PatchDiffusion. We attribute this advantage to the direct mapping between the two distributions, which mitigates information loss from noise-based initialization and avoids the uncertainty introduced during stochastic sampling in classical diffusion models.

Owing to its one-step inference, I$^2$SB achieves exceptional computational efficiency, reconstructing a slice in just 0.19\,s. Compared with cDDPM (135\,s per slice), this represents more than a 700-fold speedup (Tab.\,\ref{tab:infer_time}). This acceleration is achieved without compromising reconstruction quality, thereby improving the accessibility of extended-FOV images and enhancing their applicability in downstream clinical and research tasks. Among other diffusion models, DiffusionGAN also achieves rapid inference (0.57\,s per slice) by using large sampling time steps and delivers comparable image quality. However, the diffusion GAN still requires adversarial training, which makes its training process less stable compared to I$^2$SB.

We validated I$^2$SB and other diffusion models on both simulated noisy data and real head CBCT data. For the real data, the most prominent errors occurred in the soft tissues near the head boundary. This is primarily because the WCE method assumes that missing projection data corresponds to soft-tissue (water-equivalent) attenuation, making it challenging for deep learning models to accurately estimate the true boundary. In addition, the training data were simulated from the CQ500 dataset using the same C-arm CBCT system, introducing an inevitable domain gap between simulated training data and real test data. Future work will focus on acquiring more real CBCT data to train I$^2$SB, which is expected to further improve reconstruction quality and generalization performance.

It is worth noting that in Tab.\,\ref{tab:metrics_n} PatchDiffusion achieves reasonable RMSE and PSNR but much worse SSIM than the other methods. SSIM measures image similarity based on local statistics (luminance, contrast, and structure) computed within a sliding window. With the default small window size (7$\,\times\,$7), SSIM is highly sensitive to local structural misalignment, small contrast variations, and texture discrepancies. PatchDiffusion reconstructs images in a patch-wise manner and, while producing visually coherent and perceptually high-quality results, may introduce subtle local variations at patch boundaries or fine-scale texture differences. These variations have minimal impact on global pixel-wise fidelity (hence the reasonable RMSE and high PSNR) but are penalized disproportionately by SSIM when evaluated with a small local window.

While this study provides a systematic comparison among representative diffusion-based approaches for CT FOV extension, it does not include quantitative comparisons with several strong non-diffusion methods, such as HDeepFoV \cite{fonseca2021evaluation} or dual-domain transformer-based frameworks \cite{gao2023transformer}. Although these approaches are discussed in the related work, the lack of publicly available code or reproducible implementation details prevents a fair and controlled evaluation on our dataset. Therefore, only FBPConvNet and Pix2pixGAN were compared in this work. This limitation should be considered when interpreting the comparative performance claims. Our conclusions are therefore restricted to diffusion-based models and primarily highlight the efficiency advantage of I$^2$SB within this class. Future work will aim to include broader baseline comparisons as implementations become available or through re-implementation efforts, enabling a more comprehensive assessment across modeling paradigms.

Although diffusion-based models are often associated with multi-step sampling, the proposed I$^2$SB framework allows flexible control over the NFE at inference time. In this work, we focus on the limited-FOV CT reconstruction task and empirically observe that using NFE $=1$ provides a favorable balance between reconstruction quality and computational efficiency. As shown in Tab.~\ref{tab:nfe_eval}, increasing NFE beyond one leads to only marginal improvements in quantitative metrics and visual quality, while incurring a substantially higher inference cost. This behavior can be attributed to the training strategy of I$^2$SB, which learns the transport process over the full diffusion trajectory through random time sampling, multiple noise levels, and bidirectional Schr\"odinger Bridge coupling, even though inference is performed in a single step. We emphasize that NFE $=1$ is not claimed to be universally optimal for all imaging tasks or datasets\cite{Li2025Diffusion}; rather, it is an application-specific choice that is particularly effective for limited-FOV CT reconstruction. In comparison with DiffusionGAN, I$^2$SB achieves comparable image quality while significantly reducing inference time, which is advantageous in clinical scenarios where computational efficiency and throughput are critical.

Although the proposed I$^2$SB-based method employs one-step sampling during inference, it fundamentally differs from deterministic non-diffusion models such as FBPConvNet and Pix2PixGAN. Importantly, one-step sampling does not imply one-step training. While inference is collapsed to a single update for efficiency, the model is trained over the full diffusion trajectory by randomly sampling time steps $t$, exposing the network to multiple noise levels and enforcing consistency across the entire transport path. Through noise-conditioned denoising score matching and the bidirectional coupling inherent to the Schr\"odinger Bridge formulation (i.e., forward and backward processes), I²SB learns an entropy-regularized optimal transport between the source and target distributions rather than directly regressing a point estimate. This training paradigm imposes a smooth and globally consistent transport geometry and implicitly regularizes the solution space across multiple scales. As a result, even when evaluated with one-step inference, the learned mapping retains the benefits of diffusion-based modeling, including improved robustness to ill-posedness, reduced hallucination artifacts, and better preservation of anatomical structure. In contrast, FBPConvNet and Pix2PixGAN lack explicit distributional transport constraints and are more prone to overfitting or unstable adversarial training, particularly in limited-FOV reconstruction.

While the results on this clinical dataset are promising, it is important to note that this evaluation is based on a single test case. The primary purpose of this experiment is to demonstrate proof of concept and technical feasibility on real acquisition data with clinically realistic noise and geometry. It shows that a model trained on a large, independent simulation dataset can generalize to a real scan without fine-tuning. However, these results do not establish broad clinical validity. A rigorous clinical study with a larger, diverse patient cohort is required to fully assess the statistical performance, robustness across anatomical variations, and ultimate clinical utility of the method.

Extrapolating anatomical structures beyond the acquired CT field of view is an inherently ill-posed problem: once the data are truncated, any content outside the FOV must be inferred from prior information rather than measured signals. As a result, although the FOV-extended regions generated by our method often appear visually coherent, they cannot be assumed to represent the patient’s true anatomy. This limitation implies that FOV-extended images are not suitable for diagnostic interpretation or treatment planning that relies on precise anatomical accuracy. Instead, their use should be restricted to applications in which small anatomical deviations have limited impact, such as fiducial marker detection in CBCT-guided interventions \cite{Fan2022Fiducial}, body-composition estimation \cite{XU2023102852,liman2024diffusion}, or radiotherapy dose calculation \cite{fonseca2021evaluation,gao2023transformer,Xie2023,huijben2024generating} where prior studies have shown that FOV extension can improve workflow robustness with minimal clinical risk. Our study focuses primarily on the methodological aspects of efficient I$^2$SB–based FOV extension; validating downstream clinical impact requires disease-specific datasets and anatomically realistic radiotherapy plans, which were not available for this work. Nevertheless, FOV extension has been demonstrated to be beneficial for improving radiotherapy dose planning \cite{Xie2023,huijben2024generating}. Future efforts will therefore include dedicated evaluations of dose-calculation accuracy and other application-specific metrics to more concretely assess clinical relevance.

\section{Conclusion}
In this work, we proposed an efficient and interpretable CT FOV extension algorithm based on the I$^2$SB diffusion model. By directly learning a mapping between limited-FOV and extended-FOV CT images, I$^2$SB addresses the limitations of traditional diffusion models that rely on time-consuming, noise-based sampling. Our model offers a clear generative path from input to output, preserving anatomical consistency while improving the traceability of the reconstruction process.
Extensive experiments conducted on both simulated noisy data and real data demonstrate that I$^2$SB outperforms existing state-of-the-art methods, including cDDPM and PatchDiffusion, in terms of image quality and inference efficiency. Its fast inference time of 0.19\,s per 2D slice makes it highly suitable for time-sensitive clinical applications.
Overall, I$^2$SB provides a promising solution for FOV extension in CT imaging by combining the strengths of diffusion models with the efficiency and practicality required for real-world deployment.

\IEEEpubidadjcol

\section*{Acknowledgments}
The authors declare that they have no known competing
financial interests or personal relationships that could have
appeared to influence the work reported in this paper.

\bibliographystyle{IEEEtran}
\bibliography{refs}

\section{Appendix}
\subsection{Results of Noise-Free Data}

The results of the noise-free experiments are shown in Fig.\,\ref{fig_noise_free_comparison}. It is worth noting that residual noise remains in the generated images despite the noise-free setting in the case of cDDPM, indicating incomplete denoising during the reverse diffusion process. Both PatchDiffusion and I$^2$SB achieve comparable, high-quality reconstructions and clearly outperform cDDPM, which is consistent with the results in the noisy data experiments (see Fig.\,\ref{fig_noise_comparison}). However, I$^2$SB demonstrates better generalizability for real data (Fig.\,\ref{fig_real_data}) and a significant advantage in inference efficiency (Tab.\,\ref{tab:infer_time}).

\begin{figure*}[t]
\centering

{\includegraphics[width=0.19\textwidth]{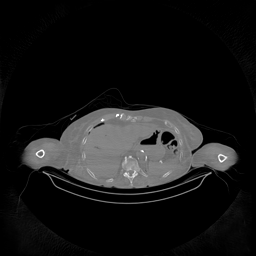}\label{fig_unet_gt}}
\hfil
{\includegraphics[width=0.19\textwidth]{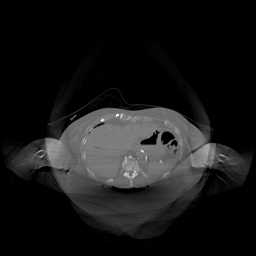}\label{fig_deeplab_gt}}
\hfil
{\includegraphics[width=0.19\textwidth]{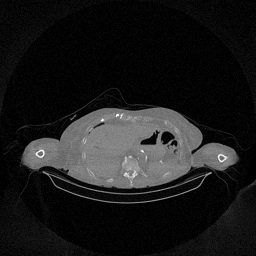}\label{fig_proposed_gt}}
\hfil
{\includegraphics[width=0.19\textwidth]{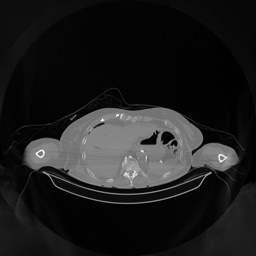}\label{fig_prunet_gt}}
\hfil
{\includegraphics[width=0.19\textwidth]{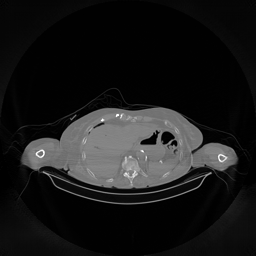}\label{fig_proposed_gt}}
\hfil

{\includegraphics[width=0.19\textwidth]{blank.png}\label{fig_unet_pred}}
\hfil
{\includegraphics[width=0.19\textwidth]{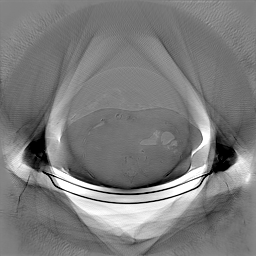}\label{fig_deeplab_pred}}
\hfil
{\includegraphics[width=0.19\textwidth]{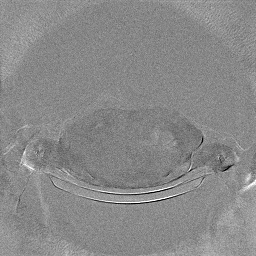}\label{fig_proposed_pred}}
\hfil
{\includegraphics[width=0.19\textwidth]{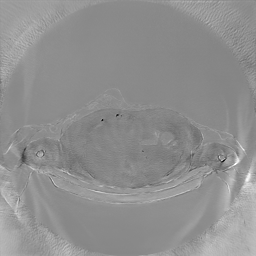}\label{fig_prunet_pred}}
\hfil
{\includegraphics[width=0.19\textwidth]{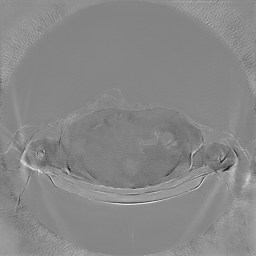}\label{fig_proposed_gt}}
\hfil

{\includegraphics[width=0.19\textwidth]{
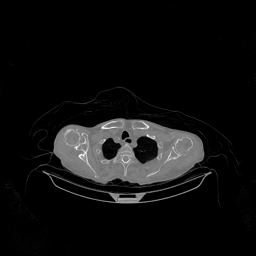}\label{fig_proposed_gt}}
\hfil
{\includegraphics[width=0.19\textwidth]{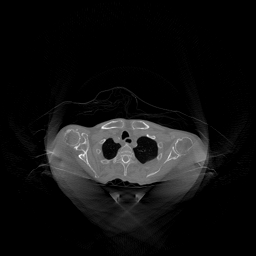}\label{fig_gan_pred}}
\hfil
{\includegraphics[width=0.19\textwidth]{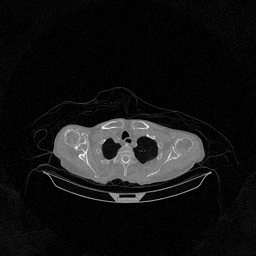}\label{fig_dosenet_pred}}
\hfil
{\includegraphics[width=0.19\textwidth]{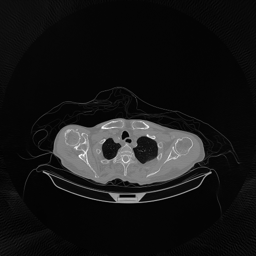}\label{fig_prunet_pred}}
\hfil
{\includegraphics[width=0.19\textwidth]{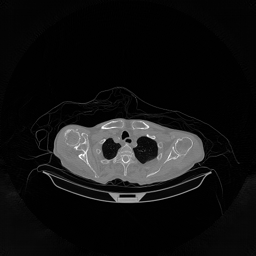}\label{fig_proposed_pred}}
\hfil

\subfloat[GT]{\includegraphics[width=0.19\textwidth]{blank.png}\label{fig_unet_error}}
\hfil
\subfloat[WCE (input)]{\includegraphics[width=0.19\textwidth]{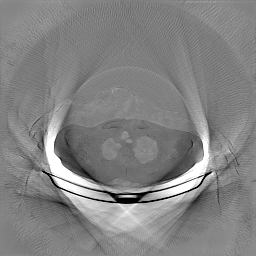}\label{fig_proposed_error}}
\hfil
\subfloat[cDDPM]{\includegraphics[width=0.19\textwidth]{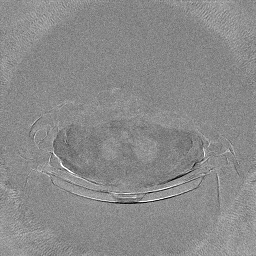}\label{fig_dosenet_error}}
\hfil
\subfloat[PatchDiffusion]{\includegraphics[width=0.19\textwidth]{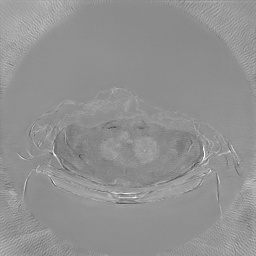}\label{fig_prunet_error}}
\hfil
\subfloat[\textbf{I$^2$SB}]{\includegraphics[width=0.19\textwidth]{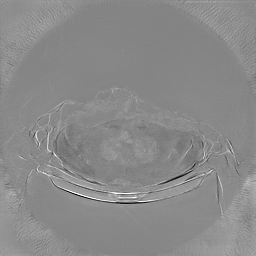}\label{fig_proposed_gt}}

\caption{Results of two exemplary test slices in the noise-free scenario. The first and third rows represent different slices in the test set, and the second and fourth rows show the error maps. The display window is [-1000, 1000]\,HU for the reconstruction images and [-500, 500]\,HU for the error maps.}
\label{fig_noise_free_comparison}
\end{figure*}

\end{document}